\documentclass{ws-p8-50x6-00}

\begin{document}

\title{Dark matter and gamma rays from the galactic halo}

\author{Philippe Jetzer}

\address{Institutes of Theoretical Physics University of Z\"urich
and ETH, Winterthurerstrasse 190, CH-8057 Z\"urich, Switzerland\\ 
E-mail: jetzer@physik.unizh.ch}

\maketitle

\abstracts{
The nature of the dark matter in the halo of our Galaxy is still
largely unknown. The microlensing events found so far towards
the Large Magellanic Cloud suggest that at most about 20\% of the
halo dark matter is in the form of MACHOs (Massive Astrophysical
Compact Halo Objects). The dark matter could also, at least 
partially, consist of cold molecular clouds (mainly $H_2$).
Another possibility is that WIMPs (Weakly Interacting Massive Particles) 
make up the dark matter and
that, due to annihilition processes, they
show up through gamma-ray emission.}

\section{Introduction}

Dixon et al.\cite{dixon} 
analysed the EGRET data concerning the diffuse $\gamma$-ray flux 
with a wavelet-based technique,
using the expected (galactic plus isotropic) emission as a null hypothesis.
Although the wavelet approach does not allow for a good estimate of the errors,
they find a statistically significant diffuse emission from 
an extended  halo surrounding the Milky Way. 
This emission 
traces a somewhat flattened halo and its 
intensity at high-galactic latitude is 
\begin{equation}
\Phi_{\gamma}(E_{\gamma}>1 {\rm GeV}) \simeq 10^{-7}-10^{-6}
~\gamma~{\rm cm^{-2}~s^{-1}~sr^{-1}}~. 
\label{eq:4}
\end{equation}
There are several possible explanations for the observed halo 
$\gamma$-ray emission. One possibility is that it is  
due to the annihilation of WIMPs 
in the halo, which would produce monoenergetic $\gamma$-rays \cite{bouquet}
 and 
also $\gamma$-rays from the decay of neutral pions generated in the particle
cascades following the annihilation.
Moreover, also positrons, neutrinos and 
in particular antiprotons would be produced, which could lead to an excess
over cosmic-ray produced antiprotons at moderately high energies
(above a few GeV).
Several authors have analysed in detail some models which predict
a $\gamma$-ray flux consistent with the one observed by EGRET.
In particular, Gondolo \cite{gondolo} considered a few GeV Majorana
fermion in a model with an extended Higgs sector. The candidate
has a mass of 2-4 GeV, a relic density $\Omega \sim 0.1$
and a scattering cross section off nucleons in the range $10^{-5}$-
$10^{-1}$ pb. The proposed model satifies present observational and
experimental constraints. 
The possibility that the $\gamma$-rays are due
to pair annihilation of neutralinos, which is the lightest supersymmetric
particle in the Minimal Supersymmetric Standard Model,
has also been extensively investigated \cite{berg}.
To explain the observed $\gamma$-ray flux a moderate amount
of clumping of the neutralinos in the halo is required, which implies also
a measurable excess of antiprotons at low energies. 
Similarly, it has been argued that neutralino annihilation
could lead to an excess of positron production, which has indeed been measured.
However, this requires a rather clumpy halo, since the annihiliation
cross section is small \cite{baltz}.
Clearly, more measurements are required to test these models.

\section{Dark clusters and cold clouds in the galactic halo}

Another possibility is that a fraction of the dark matter
is in the form of cold molecular 
clouds. 
Indeed, a few years 
ago we proposed a
scenario \cite{depaolis1,depaolis2,depaolis4,depaolisapj}, 
which relies on the Fall-Rees 
theory for the formation of globular clusters \cite{fall},
and which 
predicts that dark clusters made 
of brown dwarfs 
and cold $H_2$ clouds
\footnote{Already in 1959 Zwicky suggested that there might be
considerable amounts of $H_2$ in the Universe \cite{zw}.} should lurk in 
the galactic halo at 
galactocentric distances larger than $10-20$ kpc. Accordingly, the inner halo 
is populated by globular clusters, whereas the outer halo is dominated by 
dark clusters. 
In spite of the fact that the dark clusters resemble in many respects  
globular clusters, an important difference exists. 
Since practically no nuclear 
reactions occur in the brown dwarfs, strong stellar winds are presently 
lacking. Therefore, the leftover gas - which is ordinarily expected to
exceed 
60\% of the original amount - is not expelled from the dark clusters but 
remains confined inside them. Thus, also cold gas clouds are clumped into the 
dark clusters. Although these clouds are primarily made of $H_2$, they 
should be surrounded by an atomic layer and a photo-ionized ``skin''. Typical 
values of the cloud radius are $\sim 10^{-5}$ pc.

An important prediction of the present model is that 
high-energy cosmic-ray (CR) protons scattering on the clouds
should give rise to a detectable 
diffuse $\gamma$-ray flux from the halo of our galaxy. 
Last but not least, the issue of the origin of MACHOs 
detected since 1993 in microlensing experiments towards the 
Magellanic Clouds \cite{alcock2,laserre},
remains controversial. Although the events 
detected towards the SMC (Small Magellanic Cloud) seem to be a self-lensing 
phenomenon \cite{salati,gyuk}, a similar interpretation of all the
events 
discovered towards the LMC (Large Magellanic Cloud) 
looks unlikely~\cite{alcock2,jetzer}.
Yet -- even if most of the MACHOs are dark matter candidates lying in the 
galactic halo -- their physical nature is unclear, since their average mass 
strongly depends on the still uncertain galactic model, ranging from 
$\sim 0.1~M_{\odot}$ for a maximal disk up to $\sim 0.5~M_{\odot}$ for a 
standard isothermal sphere. 
At first glance white dwarfs look as the best explanation, 
but the resulting excessive metallicity of the halo makes this option 
untenable, unless their contribution to halo dark matter is not substantial
\cite{gm}.
So, some variations on the theme of brown dwarfs have been 
explored \cite{depaolismnras}.

Finally, we remark that ISO observations \cite{valentijn} of 
the nearby NGC891 galaxy have detected
a huge amount of molecular hydrogen, which
might account for almost all dark matter, at least within its optical radius. 

\section{$\gamma$-ray production in the galactic halo}

In the following, we estimate the halo $\gamma$-ray flux produced by the
clouds clumped into dark clusters through the interaction
with high-energy CR protons. 
CR protons scatter on cloud protons giving rise (in particular) to neutral
pions, which subsequently decay into photons. 
 
As far as the energy-dependence of the halo CRs 
is concerned, we adopt the same power-law as in the galactic disk
\cite{gaisser}
\begin{equation}
\Phi^{H}_{CR}(E) \simeq \frac{A}{{\rm GeV}} 
\left(\frac{E}{{\rm GeV}}\right)^{-\alpha}~~~
{\rm particles~cm^{-2}~s^{-1}~sr^{-1}}~. \label{eqno:42}
\end{equation}
The constant $A$ is fixed by the requirement that the integrated
energy flux agrees with the estimated value of CR energy density
in the galactic halo \cite{depaolisapjl,depaolisnjp} 
of $\rho^H_{CR} \simeq 0.12~ eV~cm^{-3}$.
Another nontrivial point concerns 
the choice of $\alpha$. As an orientation, the observed
spectrum of primary CRs on Earth would yield 
$\alpha \simeq 2.7$. However, this conclusion cannot be extrapolated
to an arbitrary region in the halo (and in the disk), since 
$\alpha$ crucially depends on the diffusion processes undergone by
CRs. For instance, the best fit to EGRET data 
in the disk towards 
the galactic centre yields \cite{mori} $\alpha \simeq 2.45$, 
thereby showing that $\alpha$ gets increased by diffusion.
In the lack of any direct information, 
we conservatively take $\alpha \simeq 2.7$ 
even in the halo,
but we checked that the flux does not vary substantially \cite{depaolisnjp}.

We now proceed with the evaluation of the $\gamma$-ray flux
produced in halo clouds
through the reactions $pp \rightarrow \pi^0 \rightarrow \gamma
\gamma$. Accordingly, the source function 
$q_{\gamma}(>E_{\gamma},\rho,l,b)$ -
yielding the photon number density at distance
$\rho$ from Earth ($l,b$ are galactic coordinates)
with energy greater than $E_{\gamma}$ - is 
\cite{gaisser}
\begin{equation}
\begin{array}{ll}
q_{\gamma}(>E_{\gamma},\rho,l,b)= 
\displaystyle{\frac{4\pi}{m_p}}  \rho_{H_2}(\rho,l,b)~\times \\ \\
\sum_{n} \int_{E_p(E_{\gamma})}^{\infty}
d\bar{E}_p  dE_{\pi}  \Phi^H_{CR}
(\bar{E}_p) 
\displaystyle{\frac{d \sigma^n_{p \rightarrow \pi}
(E_{\pi})}{dE_{\pi}}}
n_{\gamma}(\bar{E}_p)~, 
\label{eqno:46}
\end{array}
\end{equation}
where 
the lower integration limit $E_p(E_{\gamma})$ is the minimal proton 
energy necessary to produce a photon with energy $>E_{\gamma}$,
$\sigma^n_{p \rightarrow \pi}(E_\pi)$ is the cross-section for the
reaction $pp \rightarrow n \pi^0$ ($n$ is the $\pi^0$ multiplicity), 
$\rho_{H_2}(\rho,l,b)$ is the halo gas density profile
and $n_{\gamma}(\bar{E}_p)$ is 
the photon multiplicity.

Unfortunately, it would be exceedingly difficult to keep track of the
clumpiness of the actual gas distribution in the halo, and 
so we assume that its
density is smooth and goes like the dark matter density - anyhow, the very low 
angular resolution of $\gamma$-ray detectors would not permit to
distinguish between the two situations. 
Accordingly, the halo gas
density profile reads  
\begin{equation}
\rho_{H_2}(x,y,z) = f~ {\rho_0 (q)} ~ \frac{a^2+R_0^2}{a^2+x^2+y^2+(z/q)^2}~,
\label{eqno:29}
\end{equation}
for $\sqrt{ x^2+y^2+z^2/q^2} > R_{min}$, 
($R_{min} \simeq 10$ kpc is the minimal galactocentric distance of the dark clusters 
in the galactic halo). 
$f$ denotes the fraction of halo dark matter in the form of gas,
$\rho_0(q)$ is the local dark matter density, $R_0 = 8.5$ kpc is the
distance of the Sun from the galactic center, 
$a = 5.6$ kpc is the core radius and $q$ measures 
the halo flattening. For 
the standard spherical halo model 
$\rho_0(q=1) \simeq 0.3$ GeV cm$^{-3}$,
whereas it turns out that e.g. 
$\rho_0(q=0.5) \simeq 0.6$ GeV cm$^{-3}$.

At this point it is convenient to re-express 
$q_{\gamma}(>E_{\gamma},\rho,l,b)$ in terms of the inelastic pion production
cross-section $\sigma_{in}(p_{lab})$. Since
\begin{equation}
\sigma_{in}(p_{lab})<n_{\gamma}(E_p)>~ = \sum_n \int dE_{\pi}~
\frac{d\sigma^n_{p \rightarrow \pi}(E_{\pi})}{dE_{\pi}}~ n_{\gamma}(E_p)~,
\label{eqno:48}
\end{equation}
eq.(\ref{eqno:46}) becomes
\begin{equation}
\begin{array}{ll}
q_{\gamma}(>E_{\gamma},\rho,l,b)= 
\displaystyle{\frac{4\pi}{m_p}}
\rho_{H_2}(\rho,l,b) ~\times \\ \\
\int_{E_p(E_{\gamma})}^{\infty}
d\bar{E}_p~ \Phi^H_{CR}(\bar{E}_p)~ 
\sigma_{in}(p_{lab}) <n_{\gamma}(\bar{E}_p)>~, \label{eqno:49}
\end{array}
\end{equation}
where $\rho_{H_2}(\rho,l,b)$ is given by eq.(\ref{eqno:29}) with 
$x = -\rho \cos b \cos l +R_0$,
$y = -\rho \cos b \sin l$ and 
$z =  \rho \sin b$.
For the inclusive cross-section of the reaction 
$pp \rightarrow  \pi^{0}  \rightarrow \gamma \gamma$ 
we adopt the Dermer parameterization \cite{dermer}. 

Because $dV=\rho^2 d\rho d\Omega$, it follows that the observed
$\gamma$-ray flux per unit solid angle is
\begin{equation}
\Phi_{\gamma}^{~ \rm DM}
(>E_{\gamma},l,b)=\frac{1}{4\pi} 
\int^{\rho_2(l,b)}_{\rho_1(l,b)} d\rho~ q_{\gamma}(>E_{\gamma},\rho,l,b)
~. \label{eqno:51}
\end{equation}
So, we find 
\begin{equation}
\Phi_{\gamma}^{~ \rm DM}
(>E_{\gamma},l,b) = 
f ~ \frac{\rho_0(q)}{m_p}~ {I}_1(l,b)~ 
{I}_2(>E_{\gamma})~, 
\label{eqno:52}
\end{equation}
where
${I}_1(l,b)$ and ${I}_2(>E_{\gamma})$ are
defined as
\begin{equation}
{I}_1(l,b) \equiv \int^{\rho_2(l,b)}_{\rho_1(l,b)} d\rho 
\left(\frac{a^2 + R_0^2}
{a^2 + x^2 + y^2 + (z/q)^2 } \right)~, \label{eqno:A5}
\label{eqno:39}
\end{equation}
\begin{equation}
{I}_2(>E_{\gamma}) \equiv \int_{E_p(E_{\gamma})}^{\infty} 
d\bar{E}_p~\Phi^H_{CR}(\bar{E}_p)~\sigma_{in}(p_{lab})
<n_{\gamma}(\bar{E}_p)>~, 
\label{eqno:35}
\end{equation}
and $m_p$ is the proton mass.
According to our model
typical values of $\rho_1(l,b)$ and $\rho_2(l,b)$ in eqs. 
(\ref{eqno:51}) and (\ref{eqno:39})
are 10 kpc and 100 kpc, respectively.

\section{Discussion}

Regardless of the adopted value for the halo flattening parameter $q$,
$\Phi_{\gamma}^{~\rm DM}(E_{\gamma}>1{\rm ~GeV})$ lies in the range
$\simeq 6-8 \times 10^{-7}$ $\gamma$ cm$^{-2}$ s$^{-1}$ sr$^{-1}$
at high-galactic latitude. 
Thus the predicted
value for the halo $\gamma$-ray flux at high-galactic latitude
is very close to that found by Dixon et al.\cite{dixon}.
This conclusion holds almost irrespectively of the flatness parameter. 
Moreover, the comparison of the overall shape of the contour lines
with the corresponding ones of Figure 3 in Dixon et al.\cite{dixon} 
suggests that models 
with flatness parameter
$q \sim 0.8$ are in better agreement with the data, 
thereby implying that most likely 
the halo dark matter in form of $H_2$ clouds is not spherically distributed. 

Nevertheless, given the large 
uncertainties both in the data and in the model parameters
one might also explain the 
observations with a nonstandard Inverse Compton (IC)
mechanism, whereby $\gamma$-ray photons are produced by 
IC scattering of high-energy CR electrons off galactic background photons. 
Our calculation \cite{depaolisnjp}, 
however, points out that the corresponding IC
contour lines 
decrease much more rapidly than the observed ones for the halo
$\gamma$-ray emission (see Figure 3 in Dixon et al.\cite{dixon}). 
Of course, more precise measurements with a next generation
of satellites are certainly needed in order to settle the issue.

As M31 resembles our galaxy, the discovery of Dixon et al.\cite{dixon} 
naturally leads to the expectation that the halo of M31 should give rise to 
a $\gamma$-ray emission as well. 
Clearly, a good angular resolution of about one degree or less is  
necessary in order to distinguish
between the halo and disk emission from M31. 
So, the next generation of $\gamma$-ray satellites 
like GLAST will hopefully be able to test
the predictions of our model and to discriminate among the various 
proposed explanations for the $\gamma$-ray flux observed in the 
halo ranging from 
WIMP annihilation to a nonstandard IC mechanism.

\end{document}